# Employing Channel Probing to Derive End-of-Life Service Margins for Optical Spectrum Services

KAIDA KAEVAL[1,5,*], FRANK SLYNE[2], SEBASTIAN TROIA[3], EOIN KENNY[4], KLAUS GROBE[1], HELMUT GRIESSER[1], DANIEL C. KILPER[2], MARCO RUFFINI[2], JOSE-JUAN PEDRENO-MANRESA[1], SAI KIREET PATRI[1], GERT JERVAN[5]

[1] *ADVA Optical Networking SE, Martinsried, Germany*
[2] *TRINITY College Dublin, Ireland*
[3] *Politecnico di Milano, Italy*
[4] *HEAnet, Dublin, Ireland*
[5] *Tallinn University of Technology, Tallinn, Estonia*
**Corresponding author: kkaeval@adva.com*

**Optical Spectrum as a Service (OSaaS) spanning over multiple transparent optical network domains, can significantly reduce the investment and operational costs of the end-to-end service. Based on the black-link approach, these services are empowered by reconfigurable transceivers and the emerging disaggregation trend in optical transport networks. This work investigates the accuracy aspects of the channel probing method used in Generalized Signal to Noise Ratio (GSNR)-based OSaaS characterization in terrestrial brownfield systems. OSaaS service margins to accommodate impacts from enabling neighboring channels and end-of-life channel loads are experimentally derived in a systematic lab study carried out in the Open Ireland testbed. The applicability of the lab-derived margins is then verified in the HEAnet production network using a 400 GHz wide OSaaS. Finally, the probing accuracy is tested by depleting the GSNR margin through power adjustments utilizing the same 400 GHz OSaaS in the HEAnet live network. A minimum of 0.92 dB and 1.46 dB of service margin allocation is recommended to accommodate the impacts of enabling neighboring channels and end-of-life channel loads. Further 0.6 dB of GSNR margin should be allocated to compensate for probing inaccuracies.**

## 1. INTRODUCTION

Optical Spectrum as a Service (OSaaS) can provide remarkable investment and operational cost benefits [1-3]. In addition to the increased availability and decreased latency, these benefits span over five primary cost areas – reduction in capital investment from the transponder or demarcation equipment, reduction in energy consumption, which leads to reduction in $CO_2$ production, reduction in required human resources to operate and maintain the equipment, and reduction of waste products, such as utilized equipment after end-of-life. These scale linearly with the number of operator domains traversed by the end-customer signal, ramping up the interest to take these services out of the lab and implement them in production networks.

In essence, an OSaaS is a transparent lightpath connecting two endpoints in a single or multi-domain optical network [4, 5]. Differentiated from an alien wavelength service, OSaaS is capable of transporting multiple carriers over a predetermined spectrum slot. If the desired spectrum is continuously available in all included network segments, OSaaS can be operated over thousands of kilometers with transponders installed only at the end nodes of the connectivity. The performance characterization of such service may have multiple reasons, starting with Service Level Agreement documentation and achievable capacity calculations and ending with transceiver pre-emphasis.

While multiple off-line QoT estimation tools or open-access initiatives target precise performance estimations [6-9], the calculations of the actual per-wavelength performance require detailed knowledge about the underlying infrastructure and system components. This includes data about noise figures and gain profiles of the amplifiers, polarization-dependent loss and gain, filtering penalties, Optical Line System (OLS) channel allocation plan and load. This content is often unavailable or outdated for brownfield systems, originally not designed for open and disaggregated networking concepts. Furthermore, data describing the OLS design in great detail, is mostly handled as business-sensitive information and hence not shared with the end customers. To overcome this, an experimental method to characterize any open system with Optical Signal to Noise Ratio (OSNR) and Generalized Signal to Noise Ratio (GSNR) has been proposed by the SubOptic Open Cables Working Group [10]. The existing variations for these metrics alongside a comprehensive characterization guide using a channel probing, or otherwise known as Inverse Back-to-Back Method to characterize the submarine cables are explained in [11]. Further specifics about the test schema and achievable precision are discussed in [12]. However, terrestrial networks are often more prone to dynamic changes caused by changes in traffic load and exposure to temperature changes compared to submarine systems. In addition, the amplifiers are often operated in a constant gain regime and adding a probing light transceiver (PLT) together with adjacent neighboring channels may cause spectral hole burning in lightly populated systems. In addition, as the idle spectral parts in the system are not filled with Amplified Spontaneous Emission (ASE) noise loading or dummy channels, the amplifier's gain spectrum may differ between beginning-of-life and end-of-life. Furthermore, if the transceivers used for OSaaS characterization are not fully characterized against the performance impairments caused by test wavelength, chromatic dispersion, and polarization mode dispersion (PMD), it can cause bias in the estimations. Therefore, the exact procedure described in [11] and [12] is not often applicable in the characterization of OSaaS for brownfield terrestrial systems. However, as the multiple benefits of OSaaS outweigh the operational complications that the service model may introduce, the OSaaS characterization activity in terrestrial links is likely to become a part of daily work. This is because it enables precise pre-handover capacity estimations in the concatenated multi-domain network segments, operated by different operators [13].

This paper focuses on terrestrial brownfield systems, that have not been optimized for open and disaggregated networking, and offers three primary contributions to the existing knowledgebase:

1) A comprehensive overview of error sources related to channel probing method in terrestrial systems,

2) Further elaborations on our previous work [14] where, for the first time, channel probing method is used in the systematic lab study to derive OSaaS service margins required to accommodate the impact from enabled direct neighbors and the end-of-life channel loads in the OLS, and

3) using margin depletion method in the HEAnet production network to explore the changes in GSNR implementation margin required for achievable capacity estimations [15, 16] through decreasing or increasing the power levels near zero-margin, pre-Forward Error Correction (FEC) Bit Error Rate (BER) thresholds.

In addition, we further discuss the GSNR estimation inaccuracies obtained by using an available transceiver and readily available, non-transceiver-specific back-to-back characterization curve, similarly to [17] and compare the margins derived through systematic measurements in the Open Ireland testbed with the margin requirements in the HEAnet live network.

The rest of the paper is organized as follows. Section 2 provides the reader with necessary background information about the used channel probing method to estimate the GSNR of the link. Section 3 presents an overview of both the live and lab environment test setups used in this work. Section 4 presents the characterization results for four transceivers followed by the theoretical analyses for the link GSNR estimation variations caused by using a non-transceiver-specific characterization curve to interpret the Q values retrieved from the PLT. The possible variations are then experimentally demonstrated using a readily available characterization curves per transceiver type. The fifth section provides results from the systematic lab study with the goal to derive suitable OSaaS service margins to accommodate performance degradations from both neighboring channel impact and general system load increase. Section 6 compares the estimated GSNR implementation margin with the actual margin of the link. This is done by depleting or increasing the carrier power in the HEAnet network. Finally, the last section provides conclusions and takeaways from this work.

## 2. BACKGROUND

The deregulation of the telecoms market over the last 30 years has created competitive retail and wholesale markets consisting of many independent optical network providers. In order to deliver end-to-end services to customers, telecoms operators are often required to build solutions spanning multiple different network operators' networks. OSaaS is the next step in providing low latency and flexible capacity across multiple optical domains to retail and wholesale customers. Based on the ITU-T Recommendation G.807 [18], describing the generic functional architecture of the optical media network, OSaaS can be implemented in any OLS using one of the four configuration options introduced in [19] to select the width and spectral location of the Media Channel (MC) building blocks in the OLS. Since the OSaaS configuration has a high impact on the performance of the individual Optical Tributary Signals (OTSi) and hence the throughput within the provided spectrum, each OSaaS service should be characterized with both a GSNR profile and expected service margins. While the industry has accepted the channel probing method, which is also called the Inverse Back-to-Back method, as a relatively straight-forward method to experimentally evaluate the GSNR of the service, the fully characterized transceivers as per [11] are not often available for the task. This means that operators are relying on average performance data curves provided by vendors or they are obligated to capture the Q-over-OSNR curves by themselves. This requires time to go through multiple configurations and, although simple in principle, may create confusion in calculating the OSNR, when working with fixed resolution optical spectrum analyzer (OSA) and high symbol rate signal formats. Furthermore, due to manufacturing and component impairments, each of the transceivers may have a slightly different performance, suggesting the exercise should be carried out every time a new transceiver is used for the OSaaS characterization. For daily operations, this may seem like too much hassle, and operators are willing to use a good-enough readily available characterization curve with the smallest likelihood to overestimate the performance.

Based on [11], the channel probing method uses a characterized probing light transceiver to evaluate the actual channel performance of a lightpath. For this, probing-light with a fixed modulation format and symbol rate is inserted into the network in the corresponding channel location, and the pre-FEC BER estimation of the receiver, converted into a Q-value, is used to estimate the respective effective Generalized Optical Signal to Noise Ratio ($GOSNR_{\mathrm{est,link}}$). ITU-T Recommendation 977.1 [20] defines the link GOSNR as per equation (1), where both signal and noise power are referenced to the same optical bandwidth (often 0.1 nm (12.5 GHz)). The $OSNR_{\mathrm{ASE}}$ in (1) is the OSNR component caused by the ASE noise only. The $OSNR_{\mathrm{NLI}}$ and $OSNR_{\mathrm{GAWBS}}$ in (1) are the OSNR components from nonlinear distortions, and Guided Acousto-optic Wave Brillouin Scattering, respectively:

$$\frac{1}{GOSNR} = \frac{1}{OSNR_{\mathrm{ASE}}} + \frac{1}{OSNR_{\mathrm{NLI}}} + \frac{1}{OSNR_{\mathrm{GAWBS}}} \quad (1).$$

Specific to $GOSNR_{\mathrm{est,link}}$ obtained through channel probing, the individual subcomponents of the $GOSNR$ are not accessible and the estimated $GOSNR_{\mathrm{est,link}}$ value also includes impairments from transceiver implementation, narrow-band filtering, or other, that can not be separated if comprehensive PLT characterization for modem impairments has not been completed as per [11]. This means, perceived $GOSNR_{\mathrm{est,link}}$ by the transceiver is only applicable for the achievable capacity estimations for the same modem type.

To obtain the estimated GSNR of the link ($GSNR_{\mathrm{est,link}}$), the $GOSNR_{\mathrm{est,link}}$ is then normalized to the symbol rate of the PLT signal as per (2), where the 12.5 GHz is a reference bandwidth and $BW$ is the PLT bandwidth in GHz:

$$GSNR_{\mathrm{est,link}} = \frac{12.5}{BW} * GOSNR_{\mathrm{est,link}} \quad (2).$$

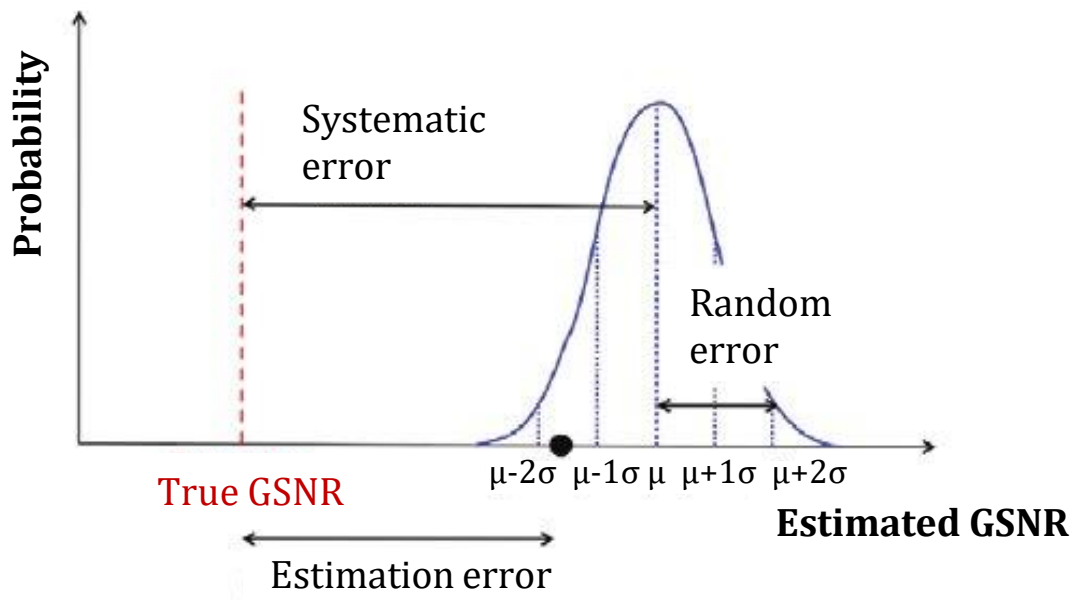

Fig 1. Error sources in GSNR estimation using channel probing

Depending on the PLT unit and network conditions during the channel probing measurements, the user must make all efforts to minimize the systematic and random error sources possibly impacting the accuracy and precision of the estimation. Knowing the impact and cause of the individual probing errors allows to reduce the estimation error magnitude and leads to a more accurate result interpretation. Fig. 1 explains the general error components in estimating the true GSNR, where μ is the estimated GSNR of the link, and σ is a standard deviation. While completely avoiding measurement errors is a hardly achievable task, the following paragraphs describe some common systematic and random error examples in terrestrial brownfield systems in order to avoid or reduce their impact on the GSNR estimation results.

The systematic error is primarily caused by inadequate PLT characterization. This can be caused by the bias from transmitter and receiver noise, wrong OSNR reading during characterization or poor polynomial fitting. The results of such errors lead to the Q-over-OSNR curve to be shifted left or right on the OSNR scale, which is directly transferred to the estimated GSNR. Also, the performance differences between the PLT and actually commissioned transceiver unit contribute to the systematic error.

If multiple recorded Q-over-OSNR curves exist for a specific transceiver type, reference [17] suggests using the average curve. Based on their example, the standard deviation (σ) of the estimated OSNR based on seven tested transceivers increased with the estimated OSNR and reached 0.7 dB at 24.05 dB OSNR. However, if minimum GSNR overestimation is desired, the best-performing curve from the transceiver characterization activity should be used. As the final estimated GSNR is achieved with the inversion, the best curve is likely to give the most modest link GSNR estimation.

To obtain meaningful performance data from the channel probing procedure, infrastructure-related error sources in channel probing should be minimized. For OSaaS use cases implemented with a different media channel width than the nominal grid of the OLS system, the systematic error with the potentially highest deviation magnitude is caused by the service equalization with nominal design-based power levels in the Reconfigurable Optical Add/Drop Multiplexers (ROADMs). To avoid this, the OSaaS services must be equalized using the maximum allowed power spectral density in the OLS. Depending on the OLS functions and capabilities, that may require some manual ROADM output power set-point adjustments to fit the OSaaS media channel width and design-based power spectral density of the OLS. The equalization is even more challenging, when several OTSi are used within a single wide-band media channel. In addition to equalization, specific spectral location under test, impact of link channel load and degradation caused by optical filtering may impact the systematic error in GSNR estimations. Furthermore, the actual performance estimations on field are dependent on the probing setup: the setup implementing a loopback at the add/drop port of the far end ROADM doubles the optical distance travelled by the probing signal. In strictly linear environments, that adds a 3 dB degradation to the $GSNR_{est,link}$ of the single direction of the link. The estimation error may be induced, if the transmit and receive directions in the network are with different performance due to very long spans at the beginning of one direction, that may deplete the OSNR early on the link. A second option for channel probing is using a transceiver pair similarly to a regular service operation in the network. However, even if both involved transceivers are characterized, it is not clear how to combine the two characterization curves of the involved transceivers for maximum probing accuracy. These specifics remain a task for future studies.

The additional bias from the actual performance can be caused also by the short-term and long-term network performance variations, as discussed in [21] that may impact the Q reading between the PLT test and the commissioning of the final transceiver unit(s), although, it can be argued if the bias caused by the delays in commissioning should be accounted as a probing error.

In general, the only random errors in channel probing can be caused by the fast network performance fluctuations. These fluctuations may be caused by PLT or OLS component, or environment instabilities that create cumulative bursts of errors, which influences the Q-value reading during the probing. Random errors contribute to reduced precision of the probing activity but can be overcome with longer measurement periods or performing multiple measurements per single PLT configuration.

As a conclusion, the channel probing method-based service characterization is only valid for the time of the measurement and various changes in the network may change the performance of the spectrum slot. In addition to PLT characterization-related estimation inaccuracies, various time-dependent variations must be accounted for. Leaving aside the fast power fluctuations and slow long-term aging, OSaaS characterization with a single probe sweep method, as proposed in [19], requires margin estimations to compensate for measurement related errors, future addition or removal of neighboring channels, and changes in general channel load.

## 3. TEST SETUP

This section introduces the test setup installed in the Open Ireland testbed and extended to HEAnet live network with the goal to characterize the ADVA TeraFlex transceiver based PLTs, investigate the GSNR estimation accuracy, derive suitable service margins in the lab and verify them in the live network.

The test setup in the Open Ireland testbed includes four important blocks:

- Noise and channel loading block, consisting of two Lumentum ROADMs 1 and 2, to generate continuous amplified spontaneous emission (ASE) load, or ASE-loaded dummy channels,
- Probing unit block, consisting of four ADVA TeraFlex transceiver based PLTs connected to the 1:8 splitter/combiner module and followed by an Erbium Doped Fiber Amplifier (EDFA) unit for loss compensation,
- Polatis series 7000 optical fiber switch that enables remote topology configurations and signal re-routing between characterization loop, lab OLS, and HEAnet live network, and
- Lab OLS, consisting of Lumentum ROADMs 3 and 4, interconnected with 25 km of G.652.D compatible fiber, and featuring a loop-back at the customer port of the far end ROADM.

In addition, an optical spectrum analyser was connected to the drop port of the 1:8 splitter/combiner, to read out the power values for characterization, and monitor the spectrum during the testing. The lab setup is illustrated in Fig. 2, left.

The lab setup was connected to the ROADM C-ports in HEAnet production network through two dedicated fiber pairs. The layout of the live network together with the interconnection site is illustrated in Fig.

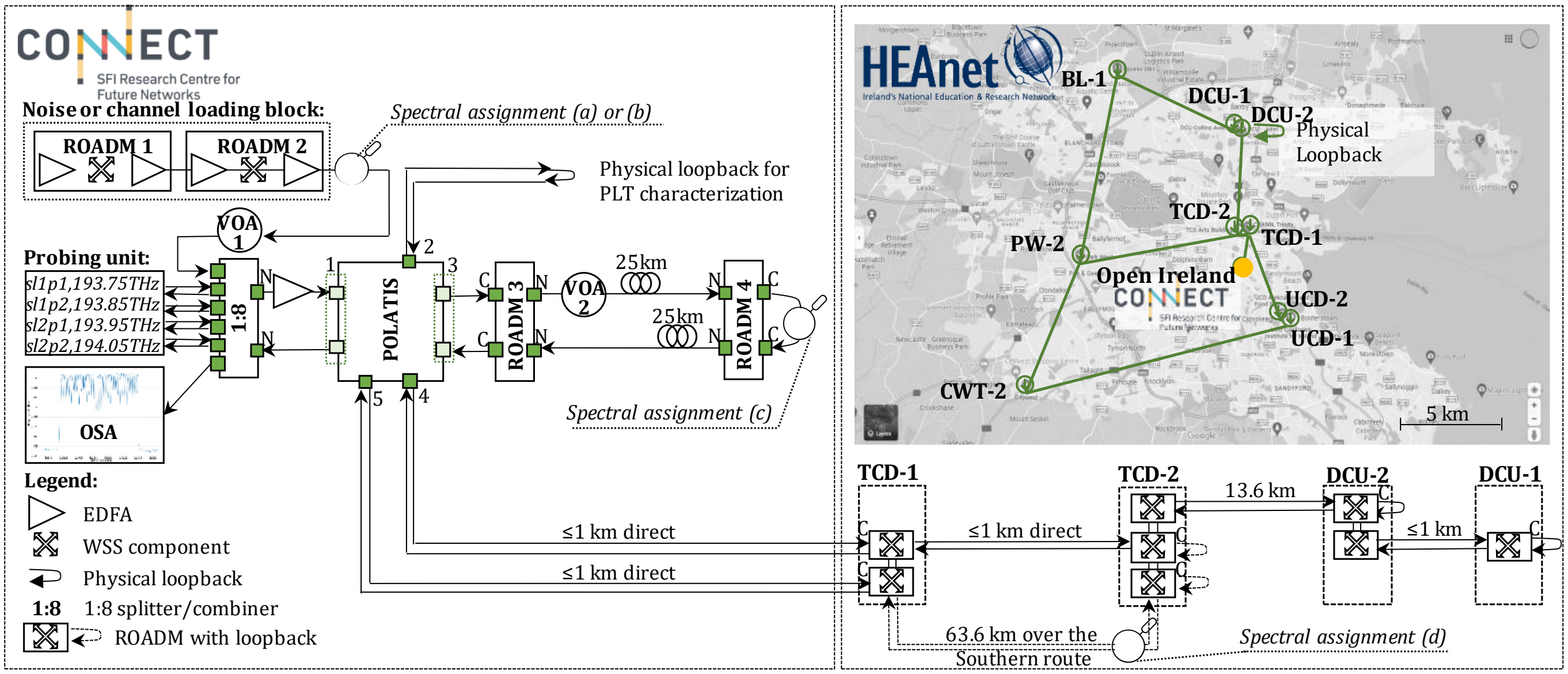

Fig. 2 General test setup for lab and live test environments

2, right. The four link lengths in the network were achieved by looping the signal back from different nodes on the two provided routes.

For the work, four PLT ports, identified by their slot location (sl) and port (p) in the probing unit, were individually characterized with Q-over-OSNR curves for 200-Gbit/s DP-QPSK 69-GBd and 200-Gbit/s DP-16QAM 34-GBd signal formats. For this, all four PLT ports were connected to the 1:8 splitter/combiner module followed by an EDFA, and individually enabled or disabled for the characterization. Continuous ASE load, covering the whole C-band, as per Fig. 3 (a) was generated in the noise- and channel loading block, by enabling the built-in amplifiers in ROADMs 1 and 2, and opening the whole spectrum for transmission in the Wavelength Selective Switch (WSS) of the ROADM 2. The power of the added noise, and hence the achievable OSNR was adjusted in variable optical attenuator (VOA)-1. The noise spectrum was then inserted to the additional 1:8 splitter/combiner module port. As all the amplifiers used in this work were operated in constant gain mode, disabling the channel under test (CuT) may change the operational state or the gain profile of the amplifiers. In order to avoid this, control channels based on available unused PLT ports, commissioned 1 THz apart from the CuT central frequency were used. For the OSNR measurement, the combined signal, consisting of noise, PLT signal under test, and two control channels, was amplified in EDFA and then connected to port 2 in Polatis switch for a loop-back. An on/off method was used to collect the power samples, disabling the channel under test to capture the noise sample at the central frequency of the CuT while leaving the control channels working.

To derive the measured OSNR, a commonly used reference bandwidth 0.1 nm (12.5 GHz) [22] was used to gather the power samples from the OSA. The measured OSNR in the PLT characterization exercise is dominated by $OSNR_{ASE}$. The OSNR was calculated as per (3), where $P_{TOT}$ was the total integrated power of the combined signal, collected over the 75-GHz signal bandwidth for 200-Gbit/s 69-GBd DP-QPSK signal and 37.5-GHz bandwidth for the 200-Gbit/s 34-GBd DP-16QAM signal, consisting of both, signal and noise component within the bandwidth. To obtain the total signal power, the noise component was subtracted from the $P_{TOT}$, using a noise power collected over a 0.1 nm (12.5 GHz) bandwidth ($N_{0.1nm}$) and adjusted by the correction coefficient based on a CuT signal bandwidth. Finally, the signal power was divided with $N_{0.1nm}$.

$$OSNR = \frac{P_{\mathrm{TOT}} - \frac{\text{CuT signal bandwidth}}{12.5} * N_{0.1\mathrm{nm}}}{N_{0.1\mathrm{nm}}} \quad (3)$$

For the experimental derivation of the required service margins to cover the impact from neighboring channels and end-of-life channel load, a dedicated OLS setup was built in the Open Ireland optical lab. It utilizes two Lumentum ROADMs with built-in boosters and pre-amplifiers, two 25 km fiber spools and two additional ROADMs from the noise and channel loading block, now providing channelized ASE-based dummy channel loading input to the connected 1:8 splitter/combiner

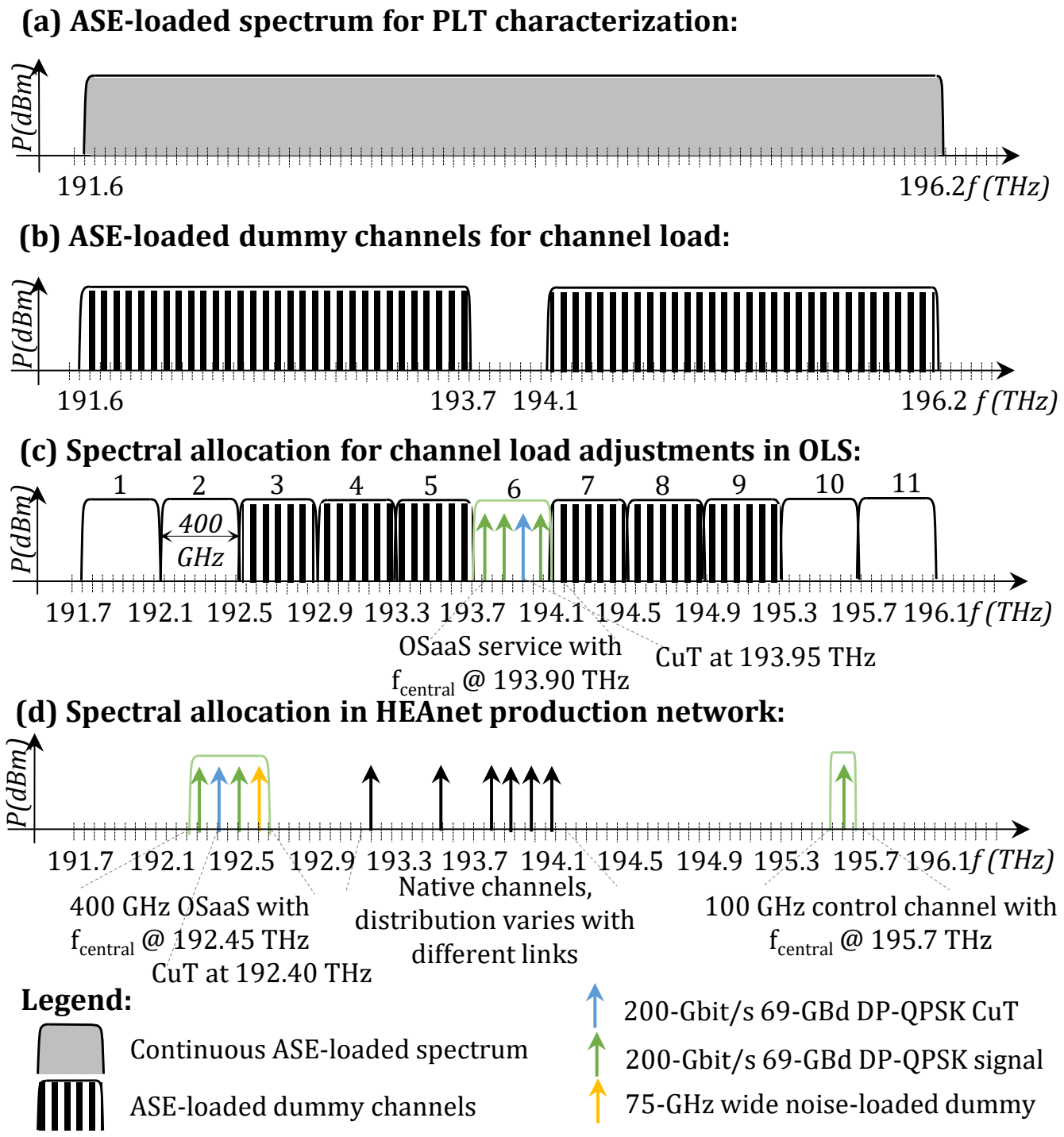

Fig. 3 Assigned spectra for (a) characterization using noise loading, (b) testing with noise-loaded dummy channels, (c) lab OLS, and (d) an example link in HEAnet production network

port, as per Fig. 3 (b). For channel loading, continuous ASE noise was generated using the booster function in the ROADM 1. Then, the ASE was shaped to 37.5 GHz bandwidth with 50 GHz spacing between each central frequency using the WSS in ROADM 2, leaving out the spectral area between 193.7 and 194.1 THz, used for OSaaS testing. All pseudo channels for channel loading were levelled at the output of the ROADM 2 and enabled or disabled by VOA-1. In lab OLS ROADMs 3 and 4, a total of eleven 400-GHz wide optical MCs, each carrying eight 37.5 GHz ASE-loaded dummy channels, were configured as per Fig. 3 (c), with the possibility to enable or disable the blocks for varied channel load conditions. The central MC block was used for channel probing. A total of nine different channel load conditions were tested. The carriers from the PLT unit commissioned in the central MC were spaced 100 GHz apart inside the MC to avoid any overlap between signal spectrums and filtering penalty on MC edges. All carriers were then configured to 200-Gbit/s DP-QPSK 69-GBd for OLS leveling. The leveling function was carried out also to ASE-loaded dummy channels to achieve a nominal 0 dBm/50 GHz Power Spectral Density (PSD) used in the lab OLS and live network. To maintain the constant PSD within the MC during testing, transmit (Tx) power of the PLT ports was individually adjusted to the symbol rate. Four different OSNR conditions were created by attenuating the power from the ROADM Tx port through the VOA-2. This reduced the input to the far-end pre-amplifier and booster and thus, reduced the effective OSNR of the whole lab OLS due to fixed gain operations of the amplifiers. Low power levels also assured, that all measurements were performed in linear operation regime in the lab.

To validate the feasibility of the derived service margins in the live network and test the GSNR margin depletion, the output power from the EDFA was connected to one of two available ROADM C-ports at the entrance node in the HEAnet's network switching the connection within Polatis switch. One 400-GHz wide OSaaS service with a 192.45 THz central frequency utilizing two unique routes and four route lengths in HEAnet's ADVA FSP3000 platform-based live production network was configured. To explain the possible loopback locations, the Northern route, spanning between TCD-1, TCD-2, DCU-2 and DCU-1 is described in detail on Fig. 2, right. The direct Southern route traversing TCD-1, UCD-2, UCD-1, CWT-2 and PW-2 to reach TCD-2 from West features no options for intermediate loopbacks. The example of the spectral allocation in live network on Fig. 3 (d). To extend the transmission distance and to allow single-ended measurements at the test site, the spectrum services were looped back at the add/drop port of the far-end ROADMs. The OSaaS in the live network was light up using four signals, transmitting with a nominal, 0 dBm/50 GHz PSD to-the-line power in the network. In addition to the three 200-Gbit/s DP-QPSK 69-GBd PLT signals, a single ASE-loaded dummy signal with a 75 GHz shaped spectrum was inserted into the HEAnet network to mimic a fourth 69-GBd DP-QPSK signal in the OSaaS bundle from the lab exercise. The fourth PLT signal was utilized as a control channel at 195.70 THz frequency, to observe any performance changes at the far end of the spectrum, when the signals within the OSaaS under test were switched on and off. The data covering these analyses is not a part of this work.

## 4. THE IMPACT OF TRANSCEIVER CHARACTERIZATION

In this section, we characterize four available PLTs with Q-over-OSNR curves to assess the GSNR estimation error caused by using a non-transceiver-specific characterization curve. 200-Gbit/s DP-QPSK 69-GBd and 200-Gbit/s DP-16QAM 34-GBd modulation format/symbol rate configurations are used for the exercise, as recommended in [11]. Then, the differences between curves are compared as per [17] and possible error in GSNR estimations is assessed when using the average curve.

Based on the transceivers available for this work, the left-hand plot in Fig. 4 presents four individual sets of transceiver-specific Q-over-OSNR datapoints, generated in Open Ireland testbed under linear conditions during characterization exercise. Although similar to the eye, the same, 14.0 dB Q-value required up to 1.38 dB higher OSNR for the worst performing PLT unit for 200-Gbit/s DP-QPSK 69-GBd signal format and achieving 12.0 dB Q-value required up to 0.37 dB higher OSNR for the 200-Gbit/s DP-16QAM 34-GBd signal format. This means that using a non-transceiver-specific curve may generate a potential GSNR estimation error. The magnitude of this error is based on the four characterized PLT transceivers used in this work is presented on the right-hand plot of Fig. 4. The y-axis presents the potential GSNR estimation error in dB. The 0-line presents the average normalized estimated GSNR from all four PLT units, and the x-axis represents the GSNR. The markers on the plot represent the maximum over- and underestimation among all four transceivers, compared to the mean estimated GSNR value calculated over all the PLT estimations. Black markers present 200-Gbit/s DP-QPSK 69-GBd, and green markers the 200-Gbit/s DP-16QAM 34-GBd signal configuration. The potential GSNR estimation error from using the non-transceiver specific curves is the difference between the curves and is dependent on the implementation penalties between the original transceiver used to create the Q-over-OSNR curve and the actual PLT unit used for probing activity. If the original PLT transceiver used for characterization curve creation was better-performing, then underestimation occurs, as the original transceivers required lower OSNR to achieve the same Q-value. If the original transceiver was performing worse than the PLT used for probing, overestimation occurs. At Q value 14.0 dB and at average estimated GSNR of 19.79 dB, the maximum variance in estimated GSNR

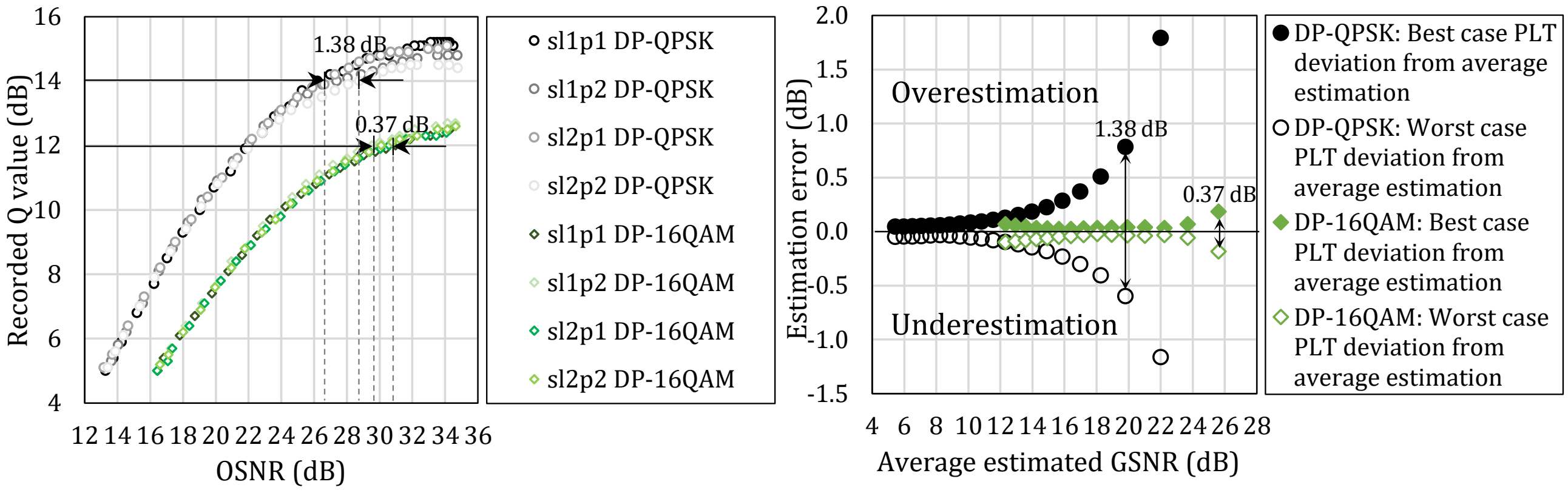

Fig. 4 Left: Characterization curves from four transceivers for 200-Gbit/s DP-QPSK 69-GBd and 200Gbit/s DP-16QAM 34-GBd signal formats
Right: Theoretical estimation error from using a non-transceiver-specific characterization curve is 1.38 dB depending on the used non-transceiver-specific curve for the 200-Gbit/s DP-QPSK 69-GBd signal. At Q value 14.5 dB and average estimated GSNR of 22.00 dB, the estimation variation between different curves grows quickly to 2.90 dB. This can make a remarkable difference in achievable capacity estimations, when Sliceable Bandwidth Variable Transceivers (S-BVT) are used. Based on [23], when keeping the symbol rate constant, a swap from 300 Gbit/s line rate to a 400 Gbit/s line rate would require 3.5 dB of extra margin, when operated with 0 ps/nm/km chromatic dispersion compensation. When applied over the full 4800 GHz of available spectrum over the C-band, this can increase the achievable throughput by 6.4 Tbit/s, when operated with 75-GHz spacing between the channels. Within one OSaaS, the gains in capacity are of course proportional to the spectral slot width, but often even 100 Gbit/s in throughput can make a big difference for operators – specifically, when the estimated GSNR was overestimated during OSaaS characterization, and the desired throughput in fact is not achievable in real life. Therefore, since high precision in detecting the GSNR of the OLS is important to accurately estimate available capacity, only transceiver-specific characterization curves are feasible at high OSNR regimes. In addition, as the curve flattening at the higher OSNR range of the Q-over-OSNR curve still degrades the accuracy of the link GSNR estimations (a0.1dB change in the Q-value can introduce a ten-fold change in the estimated GSNR), the channel performance estimations should be performed on the lower Q values of the curves and a switch to a more demanding modulation format-symbol rate combination must be performed when probing in high OSNR regimes and reaching higher GSNR estimations. This is specifically important in case of robust signal formats that have low pre-FEC bit-error-rate at high GSNRs like 100-Gbit/s DP-QPSK 31.52-GBd and 200-Gbit/s DP-QPSK 69.44-GBd.

In order to collect the data from lab OLS, the Polatis switch was reconfigured to connect the input port 1 from port 2, dedicated for characterization, to port 3, dedicated for lab OLS. First, the GSNR estimation error caused by the usage of non-transceiver-specific characterization curves was estimated. The GSNR of the lab OLS was estimated based on the Q-values collected from a single PLT port installed in sl2p1. For data collection, extended channel probing using ten PLT configurations over four different OSNR regimes was performed. For the link GSNR estimations, two transceiver-specific characterization curves from characterization exercise and a set of non-transceiver-specific, readily available characterization curves for the PLT module type were used. Fig. 5 presents the link GSNR estimation results for the transceiver-specific (TS) and non-transceiver-specific curves. The y-axis presents the absolute values for candidate GSNR estimations based on the measurements with different PLT configurations of a single PLT port. The four shaded areas present the OLS conditions after modifying the VOA-2 in the test setup, leading to four received OSNR regimes as perceived by 200-Gbit/s DP-QPSK 69-GBd in a single channel (i.e. .....I.....) load condition: 30 dB (A), 27 dB (B), 23 dB (C), and 19 dB (D), respectively. The x-axis presents nine symbols for channel load conditions, where each "I" and "." stands for enabled or disabled channelized ASE-loaded 400-GHz OSaaS service blocks from Fig. 3 (c). The GSNR estimations using the transceiver-specific curves from Fig. 4, left, exactly matching the sl2p1 actual performance, are presented with green markers, with full marker presenting 200-Gbit/s DP-QPSK 69-GBd configuration and transparent marker the 200-Gbit/s DP-16QAM 34-GBd configuration. The measurement results received with non-transceiver-specific characterization curves are presented with black and orange markers for high and low symbol rate configurations.

In Fig. 5, the minimum variation over all estimated GSNRs is at 14 dB GSNR, and the highest variation is observed near 21 dB GSNR. This also reveals that the two primary formats, 200-Gbit/s DP-QPSK and 200-Gbit/s DP-16QAM, recommended for channel probing in [11], often

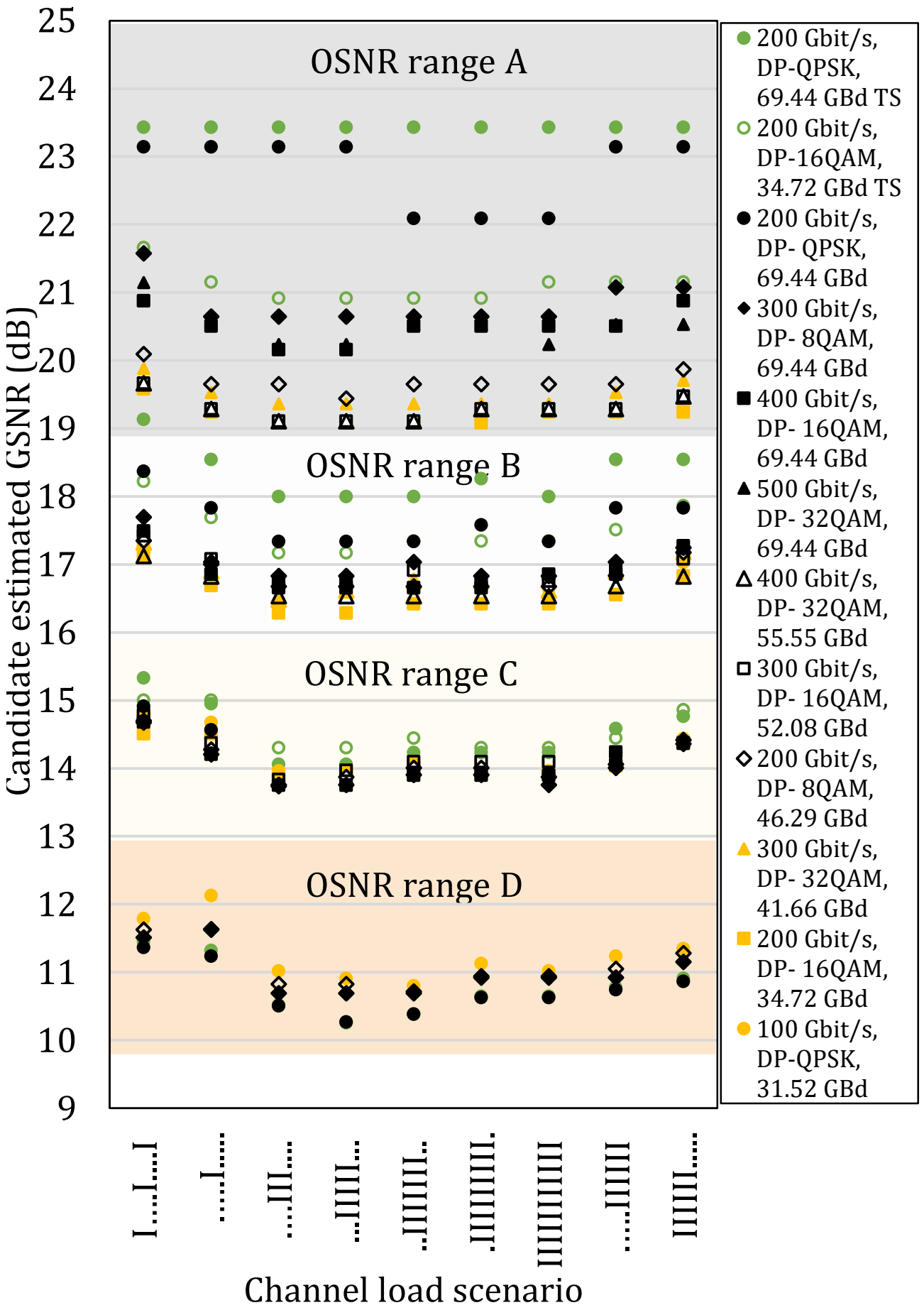

Fig. 5 GSNR estimations in different OSNR and load conditions for transceiver specific (TS) curves, marked with green markers and non-transceiver-specific characterization curves, marked with black and orange markers

present the most extremes of the estimated GSNR values from symbol-rate variable probing. The same trends between high- and low symbol rate markers can be observed for both GSNR estimations – using sl2p1 transceiver-specific curves from the characterization exercise or readily available non-transceiver-specific curves regardless of the variations in absolute GSNR estimations. As visible in Fig. 5, the transceiver-specific curves provide higher GSNR estimations at higher OSNRs compared to the non-transceiver-specific curves. As the link setup and effective OSNR is unchanged, and the Q value read out from the PLT unit remained the same, the difference in the observed performance difference between the non-transceiver-specific and transceiver-specific curves is only illusive, demonstrating a perfect example of a systematic probing error. This means the general performance of the used PLT unit in sl2p1 is slightly lower than that of the PLT unit used to create the original, non-transceiver-specific characterization curves. Although the estimations of 200-Gbit/s DP-QPSK curves are pointing to a similar GSNR range, having a maximum of 1.0 dB of estimation difference at 27 dB OSNR, the 200-Gbit/s DP-16QAM curves have a slightly higher deviation in between the estimated link GSNR, reaching up to 2.0 dB at 30 dB OSNR. As discussed, estimation errors in such magnitude can significantly impact the estimated achievable capacity in the spectrum, thus, at high OSNRs, transceiver specific characterization curves must be used whenever possible.

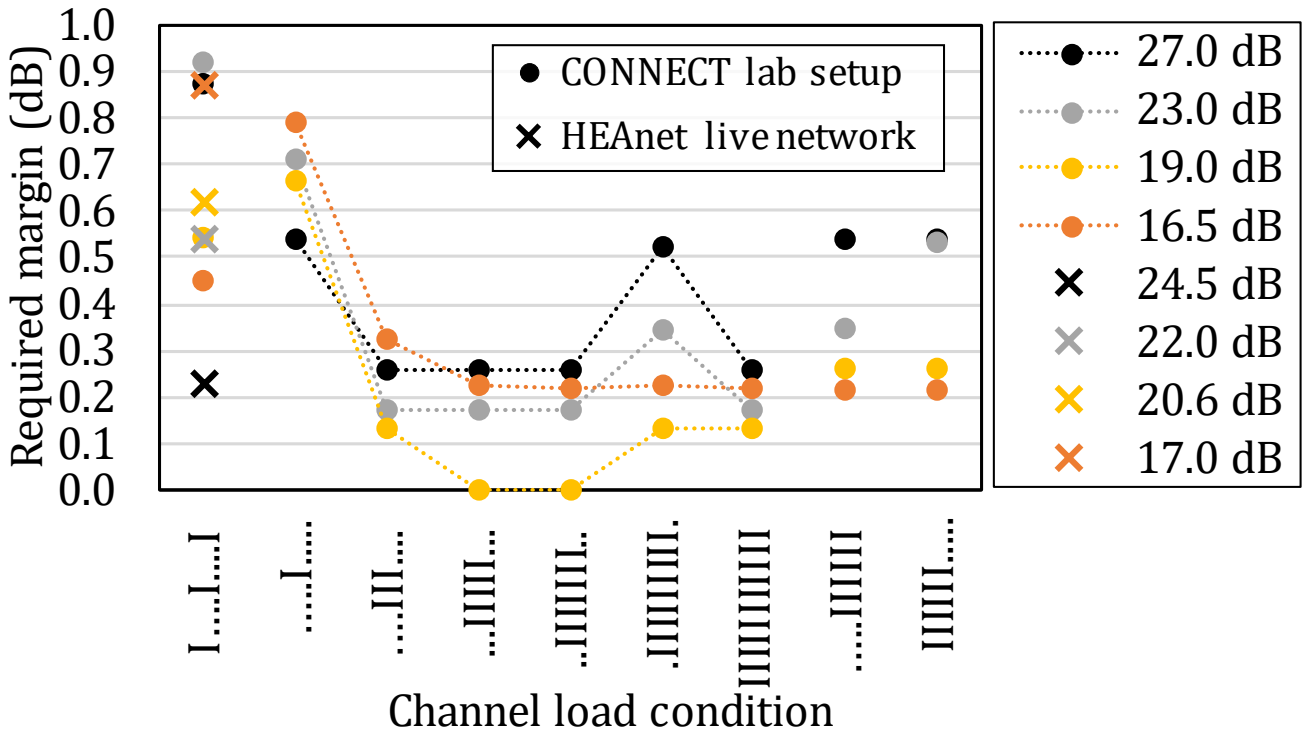

Fig. 6 Minimum required service margin to accommodate direct neighboring channels within the OSaaS for lab and live network

## 5. OSAAS SERVICE MARGINS

Reference [19] suggests, that OSaaS implementations in brownfield terrestrial systems can be characterized by using a single probe sweep procedure to capture the GSNR profile within the service spectrum. This means, that during the probing activity, the PLT is the only active channel within the provided OSaaS spectrum. This can cause some bias in estimating achievable capacity in the spectrum slot, as enabling direct neighboring channels within the OSaaS spectrum or filling in the spectrum with end-of-life channel loads can cause power redistribution and crosstalk in the spectrum and degrade the GSNR margin. In this section, we derive OSaaS service margins to cover future degradations from enabled neighboring channels and end-of-life system deployment loads. For this purpose, we focus on the 200-Gbit/s DP-QPSK 69-GBd signals, which showed the highest sensitivity to enabled neighboring channels.

Fig. 6 presents the required per-channel margin to compensate for enabling the direct neighboring channels within the OSaaS (green and yellow channels within the OSaaS spectrum, as illustrated on Fig. 3 (c) and (d)). All four PLT units within the 400-GHz OSaaS under test were configured as 200-Gbit/s DP-QPSK 69-GBd and a PLT unit installed in sl2p1, configured to 193.95 THz central frequency was used as a CuT. The shaped channels from ASE noise loading were used for end-of-life spectrum fill, which allocated 88% of the spectrum in the lab OLS, when all eleven service blocks from Fig. 3 (c) were enabled. In total, nine channel-load conditions for spectrum fill were tested, starting with lightly populated systems (I....I....I and .....I.....), followed by higher utilization density starting with (....III....) and filling the spectrum up by enabling all eleven service blocks. In addition, two additional load conditions with services enabled only in one side of the spectrum (.....IIIIII and IIIIII.....) were tested. The x-axis of Fig. 6 presents the used channel load condition and the y-axis the required GSNR margin in dB. The required GSNR margin to accommodate direct neighboring channels is calculated by subtracting the CuT performance with enabled direct neighbors from standalone performance (neighbors disabled) for every channel load condition. Line colors refer to different OSNR regimes similarly to Fig. 5. In general, the higher the spectrum allocation, the lower the impact from the addition of the direct neighboring channels. The highest required safety margin necessary to enable direct neighboring channels, based on our lab study, is 0.92 dB. This is while operating the spectrum with 100 GHz channel spacing within the OSaaS and at OSNRs below 27 dB.

In addition to the lab environment, neighboring channel impact was tested on four routes in the HEAnet live network by switching the Polatis from lab OLS to one of the ports connected with the HEAnet live network. The two routes in live network provided OSNRs between 17.0 dB and 24.5 dB. For the purposes of probing, a PLT unit installed in sl1p2 was used and reconfigured from 193.85 THz central frequency to 192.40 THz for testing in the live network. The results obtained were similar to the margin requirements derived in the systematic lab study and are included in Fig. 6 with the cross marks above scenario I....I....I, which is the closest to real-life network channel load. However, although the highest required margin in dB from production network aligns with the lab-derived margin expectations, the individual margin requirements per OSNR regime do not align for lab and live environments. This can be caused by the usage of different PLT units for the tests in lab and live environments. In real life use cases, in addition to the channel spacing, the margins can be dependent also on route length or OSNR, if the transceivers are operated in linear regime. Also, even small changes in spectral allocation, OLS components, or amplifier response may have an impact on the performance and margins. Therefore, the derived margins in this work only present one possible set of the required margins and further studies with long-term data collection from different lab and production networks are required to verify the margins that operators can commit to.

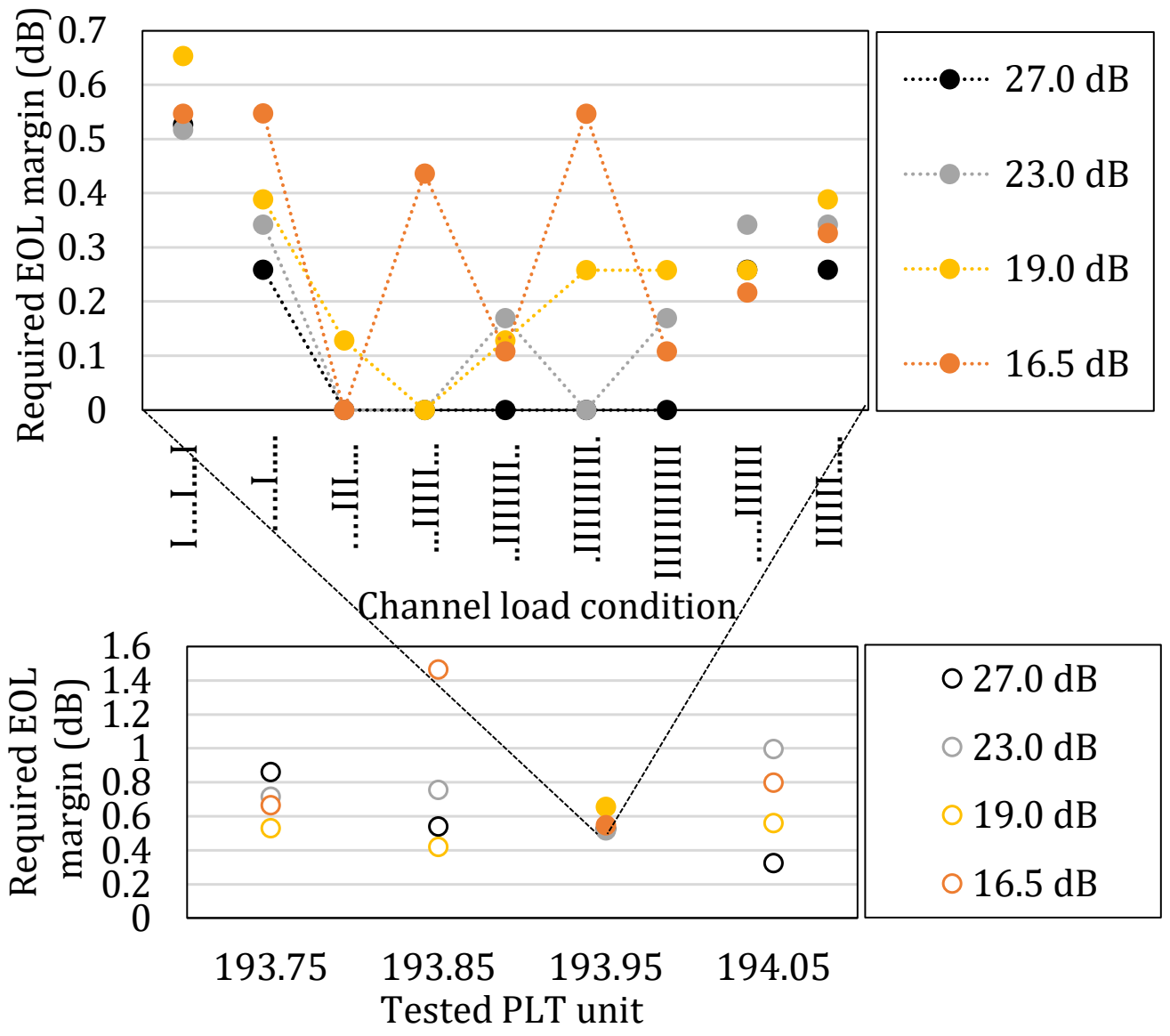

Fig. 7 Top: Minimum required EOL $GSNR_{margin}$ as per channel load conditions for PLT in sl2p1, Bottom: Minimum required EOL GSNR margin as seen by other tested PLT units

Fig. 7 illustrates the required margins to allocate end-of-life (EOL) channel loads and consists of two graphs. The top graph presents the required end-of-life GSNR margin for the CuT installed at sl2p1, with direct neighbors disabled within the OSaaS, at a certain loading condition compared to the worst CuT performance over all loading conditions. The lowest performing loading condition is preferred over maximum channel load, because the worst performance often occurred with loading conditions ....III.... to .IIIIIIIII. and not the full load. This may be due to the changes in amplifier operation regime, but the causes may be different for different OLS systems. Similarly to Fig. 6, the x-axis represents different channel load conditions and the y-axis the required margin. Line and marker styles present the OSNR conditions as on Fig. 5. Based on the results from the systematic lab study utilizing the PLT unit installed in sl2p1, 200-Gbit/s DP-QPSK 69-GBd signal configuration requires up to 0.65 dB margin allocation to cover end-of-life channel loads in the spectrum. However, different PLT units may have different margin requirements.

The bottom graph of the Fig. 7 presents only the maximum required EOL margin per OSNR regime over all loading conditions as required by different tested PLT units operated within a single 400-GHz OSaaS. The PLT units were configured with a 100-GHz spacing between the central wavelengths, starting with 193.75 THz for the sl1p1 and ending with 194.05 THz for the PLT unit installed in sl2p2. The graph is showcasing a more than two-fold difference between a relatively moderate, up to 0.65 dB EOL margin requirement for the PLT unit installed in sl2p1 (at 193.95 THz) compared to a 1.46 dB margin requirement for a PLT unit installed in sl1p2 (193.85 THz).

## 6. LIVE MARGIN ESTIMATION

In this section, we assess the GSNR implementation margin ($GSNR_{margin}$) through power adjustments in the HEAnet live network. The $GSNR_{margin}$ is used to select the best working transceiver configuration from thousands of possible configurations for the probed link and is derived directly from the estimated link GSNR. For this, typically required GSNR ($GSNR_{req}$) per configuration, available from the system specification documentation, is subtracted from the estimated link GSNR ($GSNR_{est,link}$), obtained through channel probing:

$$GSNR_{\text{margin}} = GSNR_{\text{est,link}} - GSNR_{\text{req}} \quad (4)$$

All calculations resulting in a positive GSNR implementation margin are expected to work over the probed link, and all calculations resulting in a negative GSNR implementation margin are expected not to work.

In the case of a linear operation, the GSNR margin can be viewed as a difference between the existing link OSNR and required link OSNR, as illustrated on Fig. 8. Assuming, that the measured noise power in 0.1 nm bandwidth in dBm remains the same, the margin can be increased or decreased by adjusting the power of the commissioned signal in dBm, as it effectively changes the OSNR of the CuT. When the signal bandwidth is fixed, noise conditions are constant, and the minimum required OSNR for the specific configuration is known, the reduction of the commissioned signal power to the minimum signal power required to satisfy the required OSNR should bring us close to the pre-FEC BER threshold. Therefore, the validity of the estimated $GSNR_{margin}$ can be assessed by comparing the $GSNR_{margin}$ with the relative power difference between commissioned signal power and minimum required signal power to meet the zero-margin, corresponding to pre-FEC BER threshold. For this, we first use transceiver-specific characterization curves for 200-Gbit/s DP-QPSK 69-GBd and 200-Gbit/s DP-16QAM 34-GBd PLT configurations to detect the candidate GSNR of the link. In order to avoid overestimation, the lowest estimation is used as the $GSNR_{est,link}$. Then, the required GSNR for the configuration is subtracted from the $GSNR_{est,link}$, and the estimated $GSNR_{margin}$ for various PLT configurations can be calculated as per equation (4). Configurations which have close-to-zero margins are selected as verification signals for further testing. For this, the PLT unit is configured to the configuration under interest and the power levels of the verification signal configurations were adjusted to achieve a 5.0 dB Q-value reading, exactly corresponding to the pre-FEC BER threshold and zero $GSNR_{margin}$.

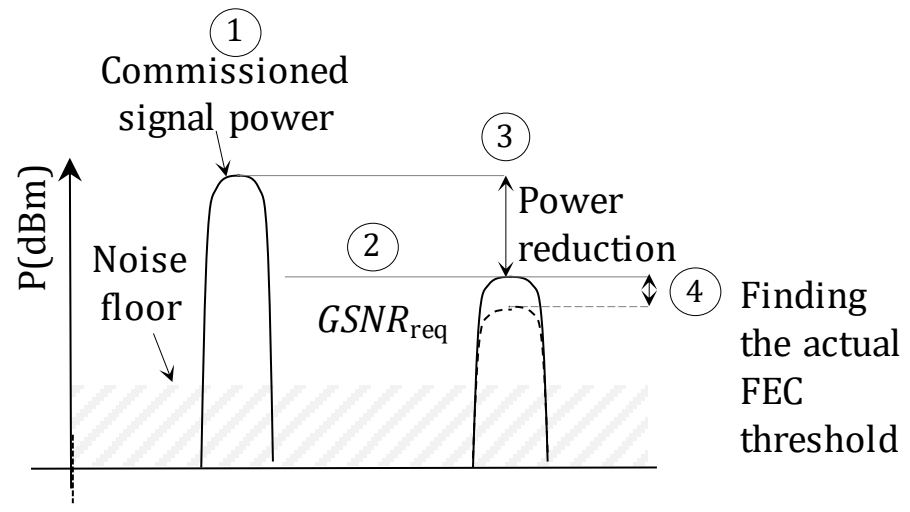

Fig. 8 Principle for margin depletion on live links

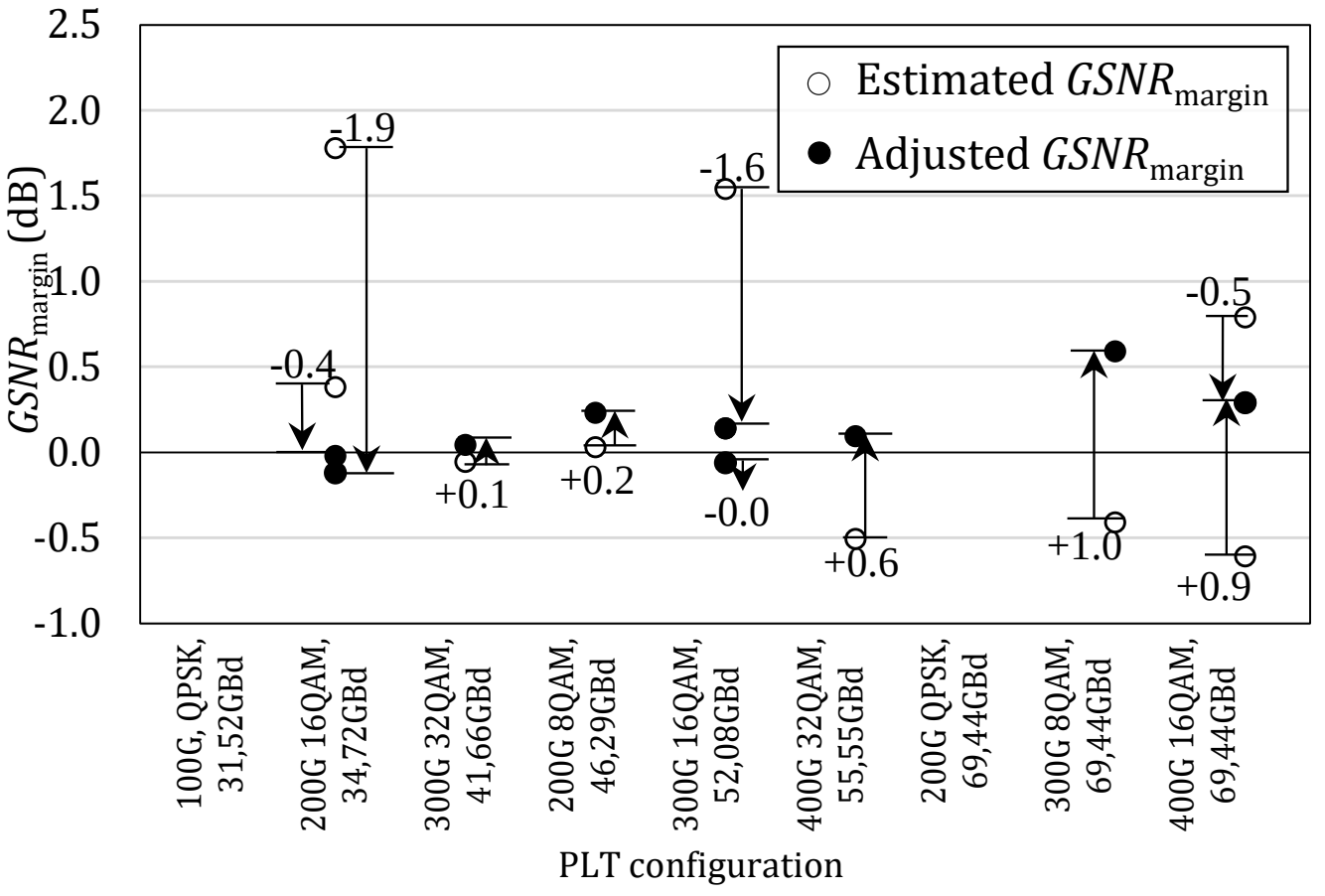

Fig. 9 Estimated and power adjusted GSNR margins from live links per configuration

Results from the margin verification activity from four live links in HEAnet network using a PLT in sl1p2 are presented in Fig. 9. The y-axis of the figure presents both, the original estimated GSNR margins, and the power adjusted $GSNR_{margin}$s with the zero-line on the plot corresponding to a zero margin, based on the $GSNR_{est,link}$. The x-axis presents all tested PLT configurations. If all the calculated margin estimations from (4) would be correct, the power adjustments in the same amount as the margin would position all the readings closely around the zero-line. Crossing the zero-margin line should change the working state of the signal. As the original state of these configurations was either below or above the FEC threshold, the arrows on the figure present the direction and amount of the power adjustment in dBm performed. If the arrow is facing up, the configuration was originally not working, and it`s power was increased. For some PLT configurations, it was possible to adjust the power levels for two live network links. In this case, the specific configuration on Fig. 9 has two estimated $GSNR_{margin}$ markers per PLT configuration. The original calculated GSNR margins as per (4) are marked with transparent markers, regardless if the channels were originally working or not. Probing results for power-adjusted verification signals are marked with black markers. Under constant noise conditions, the difference in the estimated adjustment power and required power to reach the pre-FEC BER threshold is the margin error. This can be demonstrated based on the 300-Gbit/s DP-8QAM 69.44-GBd signal format, where (4) yields GSNR margin to be -0.4 dB (i.e. the link cannot support the service with 300-Gbit/s DP-8QAM 69.44-GBd signal format). However, the probe commissioning power had to be increased by 1 dB to reach the FEC limit. The difference between this power increase and method (4) is 0.6 dB and quantifies as the inaccuracy of (4).

When the $GSNR_{est,link}$ overestimation is avoided, and lowest $GSNR_{est,link}$ is selected for the calculations, the margin verification brings us close to the zero-margin line for narrow-band signal formats. The results stay well within the probing error caused by the probing granularity (usually 0.1 dB in power adjustment increments or 0.1 dB in Q value), and only the high symbol-rate channels experience a lower margin estimation accuracy, as the margin error reaches up to 0.6 dB in error magnitude. While the most probable reason for the margin error is caused by the error in $GSNR_{est,link}$ estimations, the PLT implementation impairments, that are not characterized for the high-symbol rate PLT configurations, may also be the reason for the margin misalignments.

## 7. CONCLUSIONS

Optical Spectrum as a Service has a high potential to become one of the major building blocks for future-proofed and sustainable network topologies in Open Optical Networks. During service handover, precise service characterization together with service margins must be provided.

In this paper, we have investigated the probable GSNR estimation error during service characterization, that is caused by the usage of readily available characterization curves that are not transceiver-specific to the PLT used for the probing. We show that for estimated GSNRs below 19.79 dB, the GSNR estimation error based on four characterized PLT units can be up to 1.38 dB between the same type of PLT units for 200-Gbit/s DP-QPSK 69-GBd signal format. Any changes in the PLT internal design, that are introduced with the generation change, or replacement during the equipment repair or any other cause have a likelihood to increase this error. To benefit from the simple and straight-forward channel probing method, we recommend that the industry should look towards pre-characterized and calibrated transceivers which would be sold by the vendors as precise GSNR measurement tools. Alternatively, characterization could be automated by the vendors, and the characterization curve data can be added to the internal database of the S-BVT transceiver modules in order to enable simple and precise performance estimations by the end user.

We have further explained the experimental derivation of the margin set introduced in [14] to complement the GSNR profile-based service characterization data used during the OSaaS service handover. Based on the systematic measurements in the Open Ireland CONNECT lab, a 0.92 dB GSNR margin is required to cover degradation from enabling direct neighboring carriers, and an additional 1.46 dB margin allocation is required to compensate for the end-of-life channel load conditions in the case of 200-Gbit/s DP-QPSK 69-GBd signal configuration operation. However, as the margins are dependent on PLT configuration [14], the margins derived in this work would be applicable only for the operation with 200-Gbit/s DP-QPSK 69-GBd signal configuration. This means, further work should be performed to specify required service margins for other signal configurations that can be operated within the OSaaS.

Finally, we tested the GSNR estimation`s and GSNR implementation margin`s accuracy in HEAnet`s live network environment by adjusting the launch power to the system and capturing the state change in the signal performance. By comparing the estimated required power and actually used power to change the signal`s state, less than 0.6 dB estimation error for GSNR implementation margin was identified in a live network environment.

In summary, we find that the channel probing method to be a highly useful tool for identifying the OSaaS performance in disaggregated networking scenarios. To achieve the desired accuracy in GSNR estimations, we find that transceiver-specific characterization curves must be used. To improve the accuracy of the proposed margins and provide signal configuration independent margins that operators could commit to in their OSaaS handover documentation, further studies utilizing fully automated lab and live environment setups to gather data from various OLS setups, PLT units, and configurations should be executed. The first follow-up study in Open Ireland testbed utilizing fully automated lab OLS to estimate service margins for fully loaded OLS is captured in [24].

**Funding Information.** The work has been partially funded by the Bundesministerium für Bildung und Forschung (BMBF) under grant No. 16KIS1279K (AI-NET-PROTECT project) and Science Foundation Ireland grants 18/RI/5721 (Open Ireland), 13/RC/2077_P2 (CONNECT) and NGI Atlantic (EU grant 871582).

**Acknowledgment.** We thank HEAnet for their co-operation and help in research defining service parameters and margins for Optical Spectrum as a Service in Disaggregated Networks.

**Disclaimer.** The authors declare no conflict of interest.